\documentclass{emulateapj}

\usepackage{color}
\usepackage{lineno}
\usepackage[table]{xcolor}

\usepackage{mdframed}
\usepackage{graphicx}
\usepackage{float}
\usepackage{amsmath}
\usepackage{xcolor}

\usepackage{hyperref}
\usepackage{enumitem}
\setlist{parsep=0pt,listparindent=\parindent}
    {\pagebreak[4]\global\pdfpageattr\expandafter{\the\pdfpageattr/Rotate 90}}%
    {\pagebreak[4]\global\pdfpageattr\expandafter{\the\pdfpageattr/Rotate 0}}%

\newcommand{\DcHR}{%
  \ensuremath{\delta \langle M_B^{\mathrm{corr}}\rangle_{\mathrm{local}}}}

\newcommand{\JHU}{Department of Physics and Astronomy, The Johns Hopkins University, Baltimore, MD 21218}
\newcommand{\STScI}{Space Telescope Science Institute, Baltimore, MD 21218}
\newcommand{\Kavli}{University of Chicago, Kavli Institute for Cosmological Physics, Chicago, IL, USA}
\newcommand{\Rutgers}{Department of Physics and Astronomy, Rutgers, The State University of New Jersey, 136 Frelinghuysen Road, Piscataway, NJ 08854, USA}
\newcommand{\Moore}{Gordon and Betty Moore Foundation, 1661 Page Mill Road, Palo Alto, CA 94304, USA}
\newcommand{\Harvard}{Harvard-Smithsonian Center for Astrophysics, 60 Garden Street, Cambridge, MA 02138, USA}

\newcommand{\IfA}{Institute for Astronomy, University of Hawaii, 2680 Woodlawn Drive, Honolulu, HI 96822, USA}

\altaffiltext{1}{Department of Astronomy and Astrophysics, University of California, Santa Cruz, CA 92064, USA}
\altaffiltext{2}{\JHU}
\altaffiltext{3}{\STScI}
\altaffiltext{4}{\Kavli}
\altaffiltext{5}{Hubble, KICP Fellow}
\altaffiltext{6}{Division of Theoretical Astronomy, National Astronomical Observatory of Japan, 2-21-1 Osawa, Mitaka, Tokyo 181-8588, Japan}
\altaffiltext{7}{Institute of Astronomy and Astrophysics, Academia Sinica, Taipei 10617, Taiwan}
\altaffiltext{8}{EACOA Fellow}
\altaffiltext{9}{\Rutgers}
\altaffiltext{10}{\Harvard}
\altaffiltext{11}{\IfA}
\altaffiltext{12}{\Moore}

\def\lmstep{$0.067 \pm 0.017$}
\def\lmstepsig{3.9$\sigma$}
\def\gmstep{$0.058 \pm 0.018$}
\def\ugstep{$0.060 \pm 0.019$}
\def\ugstepsig{3.0$\sigma$}
\def\gugstep{$0.061 \pm 0.020$}

\def\flmstep{$0.091 \pm 0.024$}
\def\flmstepsig{3.8$\sigma$}
\def\fgmstep{$0.055 \pm 0.027$}
\def\fugstep{$0.120 \pm 0.030$}
\def\fugstepsig{4.0$\sigma$}
\def\fgugstep{$0.104 \pm 0.033$}

\begin{document}

\title{Should Type Ia Supernova Distances be Corrected for their Local Environments?}
\author{D. O. Jones\altaffilmark{1}, A. G. Riess\altaffilmark{2,3},
  D. M. Scolnic\altaffilmark{4,5}, Y.-C. Pan\altaffilmark{6,7,8}, E. Johnson\altaffilmark{2},
  D. A. Coulter\altaffilmark{1}, K. G. Dettman\altaffilmark{9}, M. M. Foley\altaffilmark{10},
  R. J. Foley\altaffilmark{1}, M. E. Huber\altaffilmark{11},
  S. W. Jha\altaffilmark{9}, C. D. Kilpatrick\altaffilmark{1}, R. P. Kirshner\altaffilmark{10,12},
  A. Rest\altaffilmark{2,3}, A. S. B. Schultz\altaffilmark{11}, M. R. Siebert\altaffilmark{1}}

\begin{abstract} 

 Recent analyses suggest that
 distance residuals measured from Type Ia supernovae (SNe\,Ia) are correlated with 
 local host galaxy properties within a few kpc of the SN explosion.
 However, the well-established correlation with
 global host galaxy properties is nearly as significant,
 with a shift of 0.06 mag across a low to high mass boundary (the mass step).
 Here, with 273 SNe\,Ia at $z<0.1$, we investigate whether
 stellar masses and rest-frame $u-g$ colors of regions within 1.5 kpc 
 of the SN\,Ia explosion site are significantly better correlated with
 SN distance measurements than global properties or properties measured
 at random locations in SN hosts.  At $\lesssim2\sigma$ significance,
 local properties tend to correlate with distance residuals better than properties at
 random locations, though despite using the largest low-$z$ sample to date
 we cannot definitively prove that a local correlation is more significant than
 a random correlation.
 Our data hint that SNe observed by surveys that do not target a pre-selected set of galaxies may have a larger local
 mass step than SNe from surveys that do, an increase of $0.071\pm0.036$ mag (2.0$\sigma$).
 We find a $3\sigma$ local mass step after global mass correction,
   evidence that SNe\,Ia should be corrected for their local mass, but we note that this
 effect is insignificant in the targeted low-$z$ sample.
 Only the local mass step remains
 significant at $>2\sigma$ after global mass correction,
 and we conservatively estimate a systematic shift in H$_0$ measurements
 of -0.14 $\textrm{km}\,\textrm{s}^{-1}\textrm{Mpc}^{-1}$ with an additional uncertainty of 0.14 $\textrm{km}\,\textrm{s}^{-1}\textrm{Mpc}^{-1}$, $\sim$10\% of the present uncertainty.

\end{abstract}

\section{Introduction}
\label{sec:intro}

Type Ia supernovae (SNe\,Ia) have become increasingly precise cosmological
distance indicators through improvements in how they are standardized.
Beyond accounting for the light-curve shape and color of SNe Ia,
the most recent and smallest effect to be routinely addressed in
cosmological samples is a $\sim$0.06 mag correction derived from the
empirical correlation of SN\,Ia distance residuals with host galaxy mass
(the mass step; \citealp{Kelly10,Lampeitl10,Sullivan10}).
Cosmology analyses typically correct for the mass step
\citep{Sullivan11,Betoule14} despite a lack of understanding
of the underlying cause.  If mass serves only as a proxy for
the underlying cause, for example, metallicity or progenitor age
\citep{Hayden13,Childress14,Graur15},
a somewhat different correction may yield improved cosmological
distance estimates from SNe Ia.

In cases
where the progenitor has a short delay between
formation and explosion (prompt SNe\,Ia),
the environment near the site of the SN
could be used as a better diagnostic of the properties of
the progenitor than the global host environment.
Up to 50\% of SNe\,Ia could explode less than 500 Myr after
the formation of their progenitor systems \citep{Rodney14,Maoz14}.
Therefore, a correction based on the local environment
may be a better
method of standardizing SNe\,Ia than a correction based on
the host galaxy as a whole.
However, for SNe\,Ia with longer delay times such
a correlation becomes less likely.

Evidence
for a correlation between SN shape- and color-corrected magnitude
(hereafter corrected magnitude) and local star formation rate
within 1-2 kpc of the SN explosion site
was reported by \citet{Rigault13} using SN Factory data \citep{Aldering02}
and \citet{Rigault15} for a publicly available SN sample \citep{Hicken09b}. 
\citet{Kelly15} found evidence that the dispersion of SN corrected magnitudes
was lower in highly star-forming local environments but had only a
small sample of $\sim$20 SNe that were found in such environments.
However, \citet*{Jones15b} found that
after applying updated light curve fitters and employing the
same sample selection as used for cosmological analyses,
the relationship between inferred SN\,Ia distance 
and local star formation was found to be
insignificant in a sample of 179 $z < 0.1$ SNe\,Ia.

More recently, \citet{Roman18} examined the relationship between SN corrected magnitude and local,
rest-frame $U - V$ color.  A ``step'' between blue and red colors was seen
at 1.7$\sigma$ significance at $z < 0.1$ and 6.9$\sigma$ significance
when using SNe\,Ia from $0.1 < z < 0.5$ (and 7.0$\sigma$ significance
when all SNe are included).
The $z < 0.1$ step measurement is $0.053\pm0.032$ mag, while the $0.1 < z < 0.5$
step is $0.117 \pm 0.017$ mag.  The reason for this difference is unclear.  Factors
could include statistical fluctuation,
survey selection effects, different effective apertures due to blending at high-$z$,
or a redshift-dependent local step.  Similarly, \citet{Kim18} used global properties to \textit{infer}
local properties for a subset of SNe\,Ia from $0.01 \lesssim z \lesssim 1$,
finding that the inferred local star formation correction was $0.081 \pm 0.018$ mag, 0.024 mag larger
than the global mass step.  \citet{Rigault18} recently measured a $0.163 \pm 0.029$ mag correlation
between local \textit{specific} star formation rate (sSFR) and Hubble residual using SNFactory data
but do not measure a global sSFR step.  \citet{Uddin17} also examined 1338 SN\,Ia and found that SNe\,Ia in host
galaxies with high \textit{global} sSFR had the lowest intrinsic
dispersion of the subsamples they examined.

Here, we ask whether the evidence for a local step implies
that host galaxy properties near the SN location
contain additional information that could improve the
standardization of SNe\,Ia.  Alternatively, it may be that local
regions merely trace global host galaxy properties.
\citet{Roman18}, for example, found that the size of the local step decreases by just
0.022 mag when inferring local properties within
an aperture of radius 16 kpc, an aperture that should
contain nearly all the light from a galaxy.
With the first data release of the Foundation low-$z$ SN sample
\citep{Foley18}, we are now able to ask this question with up to 273
$z < 0.1$ SNe\,Ia, a low-$z$ sample that is $\sim$40\% larger than that
used in previous cosmological analyses \citep{Scolnic18,Jones18,Betoule14}.

We use measurements of the stellar
mass and local, rest-frame $u - g$ colors near the SN location.
The local stellar mass is a natural first measurement to investigate,
given the known correlation of SN distance residuals
with global stellar mass.  Measuring local stellar mass is also
a convenient measurement; it only requires optical photometry, which is
available for the entire low-$z$ SN sample.
Rest-frame $u - g$ colors, on the other hand, are effectively the same as the local
$U-V$ colors used by \citet{Roman18}.  $u-g$ colors are sensitive to the
host galaxy star formation without suffering from the resolution
limitations of shorter-wavelength UV instruments such as GALEX
(e.g. \citealp{Jones15b}).

We measure the Hubble residual ``step''
as a function of global properties, local properties, and the properties
within 1.5 kpc apertures at random locations within each host galaxy.
In \S\ref{sec:data} we present the SN sample and we measure
host galaxy properties in \S\ref{sec:analysis}.
In \S\ref{sec:local} we
measure the correlation of these data with host galaxy properties,
and in \S\ref{sec:cosmoparam} we examine the impact of our results
on the Hubble constant.  We conclude in \S\ref{sec:conclusions}.

\section{Data and Analysis}
\label{sec:data}

For this analysis, we combine 216 $z < 0.1$ SNe from the
Pantheon compilation \citep{Scolnic18} with
178 SNe\,Ia from the Foundation first data
release (DR1; \citealp{Foley18}).  This combined sample contains
394 SNe\,Ia and twice as many SNe\,Ia
at $z < 0.1$ as recent cosmological analyses (e.g. \citealp{Scolnic18}).

The Pantheon compilation includes SNe observed by CfA surveys 1-4
\citep{Riess99,Jha06,Hicken09a,Hicken09b,Hicken12} and the
Carnegie Supernova Project \citep{Contreras10,Folatelli10,Stritzinger11}.
It also includes 43 $z < 0.1$ SNe discovered by SDSS \citep{Kessler09} and
PS1 \citep{Scolnic18,Rest14,Scolnic14}.

The Foundation survey uses the Pan-STARRS1 (PS1) telescope to follow nearby SNe\,Ia
discovered by ASAS-SN \citep{Holoien17}, ATLAS \citep{Tonry18},
Gaia \citep{Prusti16}, and the Pan-STARRS Survey for Transients
(PSST; \citealp{Huber15}) among other surveys.
SNe\,Ia from the Foundation DR1 are observed on the well-calibrated
PS1 photometric system \citep{Schlafly12} and can therefore
be used to measure distances with
good control over systematic uncertainties.
The Foundation DR1 includes 225 SNe\,Ia, 180 of which pass
the cuts for inclusion in a cosmological analysis used
in \citet{Foley18} (2 Foundation SNe\,Ia are at $z > 0.1$
and are therefore excluded here).

\subsection{Sample Selection Requirements using SALT2}
\label{sec:dist}

We infer distances from the SNe\,Ia in the Pantheon and Foundation samples
using the most recent version of the SALT2 light curve fitter \citep{Guy07}
(SALT2.4; \citealp{Betoule14,Guy10}) and the Tripp estimator \citep{Tripp98}:

\begin{equation}
\mu = m_B - \mathcal{M} + \alpha \times x_1 - \beta \times c.
\label{eqn:salt2}
\end{equation}

\noindent $m_B$ is the log of the light curve amplitude,
$x_1$ is the light curve shape parameter, and $c$ is the light
curve color parameter.  $\alpha$ and $\beta$ are nuisance parameters
along with $\mathcal{M}$, a parameter encompassing the SN\,Ia
absolute magnitude at peak and the Hubble constant.

The Pantheon and Foundation analyses apply sample selection criteria using
these SALT2 light curve parameters to ensure that the SN\,Ia included can
yield accurate distances.
These include cuts on the shape and color to ensure that
the SNe are within the parameter ranges for which the SALT2 model
is valid ($-3 < x_1 < 3$, $-0.3 < c < 0.3$), and cuts to ensure
that the shape and time of maximum light are well-measured ($x_1$
uncertainty $<$1 day and time of maximum uncertainty $<$2 days).
Here, we also require Milky Way reddening of $E(B - V) < 0.15$ mag and $z > 0.01$
to remove SNe with large systematic peculiar velocity uncertainties.

The Foundation data have a few additional selection criteria,
all of which were applied in \citet{Foley18}: the first
light curve point must have a phase of $<$7 days, at least 11 total
light curve points are required in $gri_{P1}$, and Chauvenet's criterion
is applied to remove outliers.  All samples remove spectroscopically peculiar
SNe\,Ia (apart from 1991T-like SNe, which are included).

Finally, survey selection effects bias the SN distances,
the light curve shapes, and the light curve colors.  We apply bias corrections to the distances and
light curve parameters using the BEAMS with Bias Corrections (BBC)
method \citep{Kessler16}.  The BBC method uses simulated SN
samples to correct $x_1$, $c$, $m_B$, $\alpha$, and $\beta$
for observational biases and selection effects.  Though
the BBC method makes no corrections based on host galaxy information
directly, the BBC
corrections are important for this study
because SN\,Ia light curve demographics depend on host
properties \citep{Childress13} and SNe\,Ia with $c < -0.2$ and $x_1 > 2$ have
mean Hubble residuals, before BBC correction, of up to
0.2-0.3 mag \citep{Scolnic16}.  These residuals are 3-4 times larger than the
host mass step.

We use the simulations
from \citet{Scolnic18} and \citet{Foley18} with the BBC method
to generate these bias corrections (Scolnic et al. 2018, in prep, will contain
additional simulation details specific to the Foundation sample).
The BBC method removes 6 additional SNe from the sample;
3 from Pantheon and 3 from Foundation.  We
cannot be certain that the bias corrections are valid for
these 6 SNe as they
lie in a region of shape, color and redshift space that
is not well sampled by the SN simulation.
With the BBC method,
we find $\alpha = 0.141$ and $\beta = 3.149$ using 
the z $<$ 0.1 SNe in this analysis.

After these additional cosmology cuts, 170 $z < 0.1$ SNe\,Ia are
from the CSP or CfA surveys, 43 are from SDSS or PS1,
and 170 are from the Foundation DR1 sample for a total of 383 SNe\,Ia.  We note that
\citet{Foley18} lists 180 SNe as passing all cosmology cuts.
Of these, 3 are at $z < 0.01$, 2 are at $z > 0.1$, 2 do not
pass cuts due to small changes in the SALT2 fitting
parameters\footnote{In order to match the Pantheon analysis,
  we reduce the wavelength range over which
  the SALT2 model is fit to the photometric data to
  a maximum of 7000 \AA.} and
the remainder are lost due to BBC cuts.  See \citet{Foley18} and
\citet{Scolnic18} (and references therein) for additional details on
the sample selection.

\subsection{Sample Selection Requirements using Host Galaxy Properties}
\label{sec:dist}

We measure host galaxy properties with photometry from the PS1 first data release
\citep{Chambers16} and the Sloan Digital Sky
Survey Data Release 14 (SDSS DR14; \citealp{Abolfathi17}).
The PS1 DR1 has deep, $grizy$
observations over 3$\pi$ steradians of the sky and
has observed at the locations of over 90\% of SNe
in the current low-$z$ sample.
PS1 $y$ band photometry in particular allows
for a robust determination of host galaxy masses.
SDSS has imaged $\sim$14,000 square degrees in the
$ugriz$ filters, including the locations of $\sim$65\% of the SNe in
the Pantheon+Foundation low-$z$ sample.
We measure SDSS $u$ and PS1 $grizy$ photometry within apertures
of 3 kpc diameter at the location of each SN in this sample.

To observe only the regions within $\sim$3 kpc
of the SN, we require the typical seeing of PS1 and SDSS
to correspond to 3 kpc in physical size or less.
PS1 images have a typical seeing of $\sim$1$''$, while SDSS images have a median
seeing of approximately 1.38$''$ in $u$.  Blending of local and global effects
may occur at higher redshifts.  If we therefore restrict our sample
to $z < 0.1$, where 3 kpc corresponds to an angle of
$\sim$1.6$''$, we can be assured that we are indeed probing
local regions.

We also remove 29 SNe in galaxies with inclination angles $>70^{\circ}$
based on the \citet{Tully77} axial ratio method, leaving 354.  This cut
increases the likelihood that local regions are truly
local, as highly inclined galaxies could have non-local regions
contained in the 3 kpc aperture due to projection effects.
However, we note that projection effects will always be a concern in
this type of study, particularly in early-type galaxies.
Finally, we remove SNe for which the identification of the
host galaxy is uncertain.  SNe for which the host cannot be reliably identified
should not be used in a sample that compares local
to global measurements.  We match SNe\,Ia to candidate host galaxies
using the galaxy size- and orientation-weighted SN
separation parameter, $R$ \citep{Sullivan06}:

\begin{equation}
  \begin{split}
  R^2 &= C_{xx}x_r^2 + C_{yy}y_r^2 + C_{xy}x_ry_r\\
  C_{xx} &= \cos^2(\theta)/r_A^2 + \sin^2(\theta)/r_B^2\\
  C_{xy} &= 2\cos(\theta)\sin(\theta)(1/r_A^2 + 1/r_B^2)\\
  C_{yy} &= \sin^2(\theta)/r_A^2 + \cos^2(\theta)/r_B^2,
  \end{split}
  \end{equation}

\noindent where $x_r = x_{SN} - x_{gal}$ and
$y_r = y_{SN} - y_{gal}$.  $r_A$, $r_B$, and $\theta$
are galaxy ellipse parameters measured by SExtractor
\citep{Bertin96}.
Each $R$ parameter corresponds to an
elliptical radius about the host center.
We consider the host ambiguous if the minimum
$R$ is greater than 5.  This cut removes an additional 83 SNe,
leaving a final sample of 273 SNe.

After all cuts, $grizy$ images for measuring
the local mass step are available for 273 SNe\,Ia.  195 of these SNe
lie in the SDSS footprint and therefore
have $u$ measurements for
measuring the rest-frame $u - g$ color.  We do not attempt to infer
rest-frame $u$ colors for host galaxies without $u$
observations.

\section{Measuring Host Galaxy Properties and the Hubble Residual Steps}
\label{sec:analysis}

\begin{figure*}
\includegraphics[width=7in]{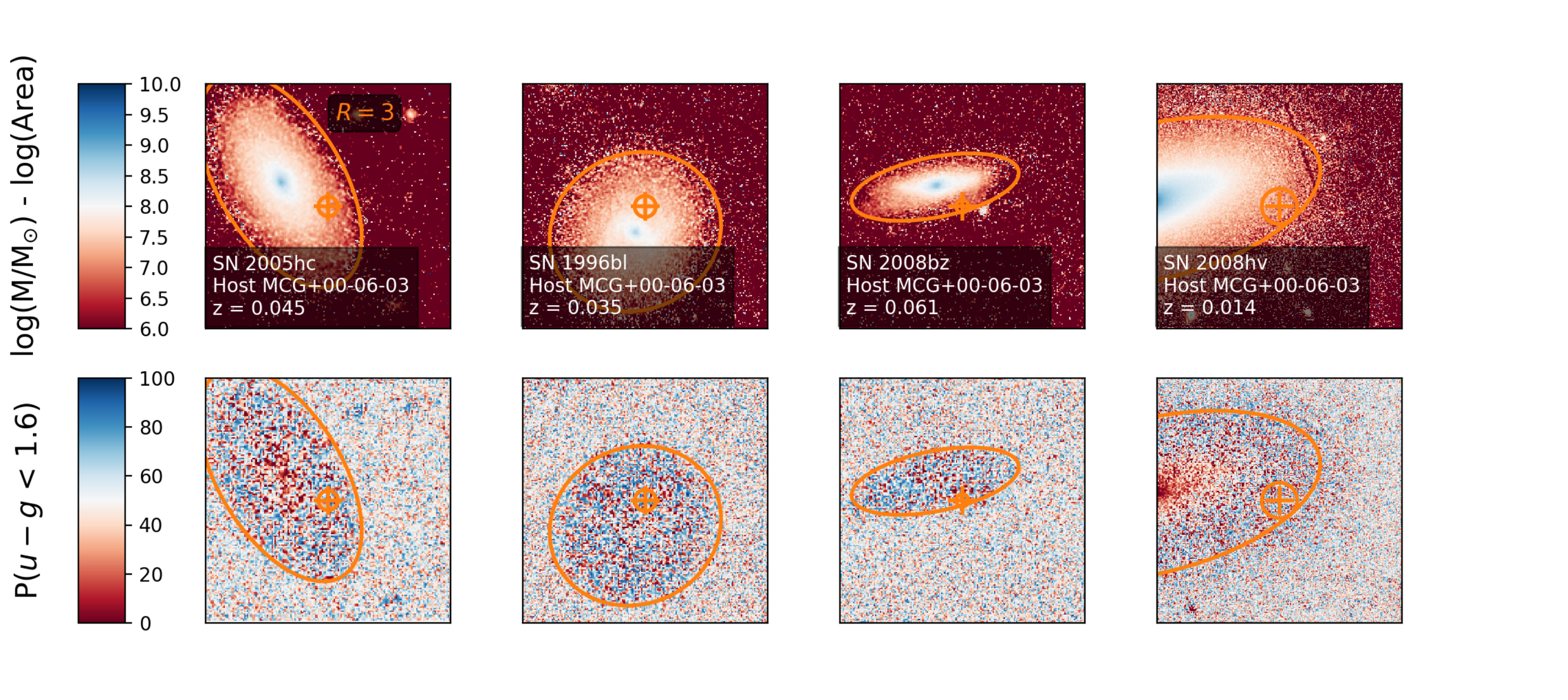}
\caption{Local mass density and $u - g$ maps from four representative galaxies in
  our sample.  The local mass and colors used in this work are measured from the 3 kpc diameter
  regions indicated by the small circles.
  For illustration, the local mass density is computed per pixel
  and has a median value of log(M$_{\ast}$/M$_{\odot})$ - log(Area) $\sim 8$ kpc$^{-2}$.
  To include regions of negative flux in the map, which have
  an undefined color measurement, the bottom row
  shows the probability that the true $u - g$ color is $< 1.6$ mag (the median observed
  color of this sample).  The approximate $R=3$ isophotal
  radius of each galaxy is denoted by the ellipses.  White colors in the map
  indicate regions on the border between locally high-mass
  and low-mass and blue $u - g$/red $u - g$ (and may also
  indicate pixels with higher than average noise).  For the purposes
  of this plot, we use observer-frame $u - g$ colors that have not
  been corrected for host galaxy reddening and Equation 8 from
  \citet{Taylor11} to approximate the host galaxy mass using
  the observed $gi$ photometry.}
\label{fig:img}
\end{figure*}

The local photometry was measured within a circular aperture of radius
1.5 kpc, while the global host galaxy photometry was measured using
elliptical aperture photometry.  The size of the global host
ellipse was set to be equal to the $R = 4$ ellipse measured by SExtractor
on each PS1 $r$-band image.  A uniform ellipse radius that extends just
beyond the estimated isophotal radius of the galaxy
ensures that all flux is captured and that a uniform aperture size
is used for all photometric bands.  An $R = 4$ ellipse is
still small enough for contamination from neighboring stars or galaxies
to be negligible.  In addition, the difference between the PS1 and SDSS seeing is
just 1.7\% of the median $R = 4$ semimajor axis of the galaxies in this sample
and therefore should not significantly bias the photometry, especially given that
the elliptical aperture extends beyond each galaxy's isophotal radius.

We then fit the local and global $ugrizy$ photometry to template SEDs following the method
of \citet{Pan14}.  We estimate galaxy masses and
un-reddened, rest-frame $u$ and $g$ colors using the \texttt{Z-PEG} SED-fitting code
\citep{LeBorgne02}, which is based on spectral synthesis
from \texttt{PEGASE.2} \citep{Fioc97}.  Galaxy SED templates
correspond to spectral types SB, Im, Sd, Sc, Sbc, Sa, S0 and E.
We marginalize over E(B-V), which is allowed to vary from 0 to 0.2 mag,
and the star formation rate.  In addition to varying E(B-V), \texttt{Z-PEG} uses 15 star formation
histories, 200 stellar age bins, and 6 metallicity bins to fit the observed photometry
and densely sample the parameter space.

Uncertainties are estimated by generating Monte Carlo realizations
of our photometric measurements.  For each filter, we generate mock photometric points from
a normal distribution with standard deviation equal to the
photometric uncertainties, and use \texttt{Z-PEG} to fit SEDs to each
realization of the photometric data.  We then estimate the uncertainty
in the host mass and rest-frame photometry from the spread in
output values.  The photometric uncertainties from this approach
can occasionally be unrealistically small; for this reason we add 0.05 mag uncertainty in quadrature
to the $u - g$ rest-frame colors, approximately equal to the photometric errors
for a 3 kpc region in a bright host galaxy,
to account for systematic uncertainties in the SED fitting.

Several studies have discussed whether local and global SED-fitting
measurements are self-consistent.  \citet{Sorba15} found a 0.1 dex bias
in global host galaxy mass measurements of star-forming galaxies when fitting
mass to the photometry of the entire galaxy instead of performing a
pixel-by-pixel fit and summing the individual measurements.  This level
of bias will not affect our results, as we look at global
and local mass independently (defining the step location separately
for global and local measurements).  The location of the step is also not
known to within 0.1 dex \citep{Scolnic18}.  Other studies have
found that summing the results of pixel-by-pixel SED fitting give
the same parameters as a SED fit to the photometry of the whole
galaxy \citep{Salim16,SanRoman18}.

\subsection{Measuring the Mass and Color Steps}
\label{sec:likemod}

We treat the dependence of SN\,Ia shape- and color-corrected
magnitude on host mass and $u - g$ as a step function,
as previous studies have found this to be well-motivated by
the data \citep{Betoule14,Roman18}.  There may be theoretical
reasons to favor a step function as well; \citet{Childress14}
predict that the mean ages of SN\,Ia progenitors undergo a sharp
transition between low-mass and high-mass galaxies.  If Hubble
residuals depend on physics related to progenitor age,
a step would naturally be produced in this model.  The dust
extinction law in passive versus star-forming galaxies could also change in
a way that would produce a step.

To estimate the size of the mass and $u - g$ color steps,
we use the maximum likelihood approach from
\citet*{Jones15b}.
Our likelihood model treats SNe in low-mass and high-mass regions
(or in regions with blue/red $u-g$ colors) as belonging
to two separate Gaussian distributions and simultaneously determines
the maximum likelihood means and standard deviations of those two
distributions.  The four parameters of this model can be
easily constrained with a standard minimization algorithm.
The baseline approach considered here does not re-fit
  $\alpha$ and $\beta$ on each side of the color or mass split,
  but we explore this approach in \S\ref{sec:nuisance}.

The step between low-mass/high-mass and bluer/redder $u-g$
may correspond roughly to the boundary between passive
and star-forming galaxies.  The median rest-frame, host galaxy
dust-corrected $u - g$ color
of this sample is 1.27, and we adopt this value as an agnostic
choice for the location of the step following \citet{Roman18}.
For the local mass step,
we again choose the divide between the locally ``low'' and ``high'' mass galaxies to
be the median local mass of our sample, log(M$_{\ast}$/M$_{\odot}) = 8.83$.
The local mass, as defined here, is the
stellar mass in the cylinder within a circular aperture of diameter 3 kpc.
Unlike the local mass or color steps, the location of
the global host mass step has been well-measured by multiple independent
datasets and analyses.  For this reason, we adopt the standard global
host mass step location of log(M$_{\ast}$/M$_{\odot}) = 10$
\citep{Sullivan10,Betoule14,Scolnic18}.
Figure \ref{fig:img} shows local
mass per pixel and $u - g$ maps for four representative galaxies in
our sample.

\begin{figure*}
\includegraphics[width=7in]{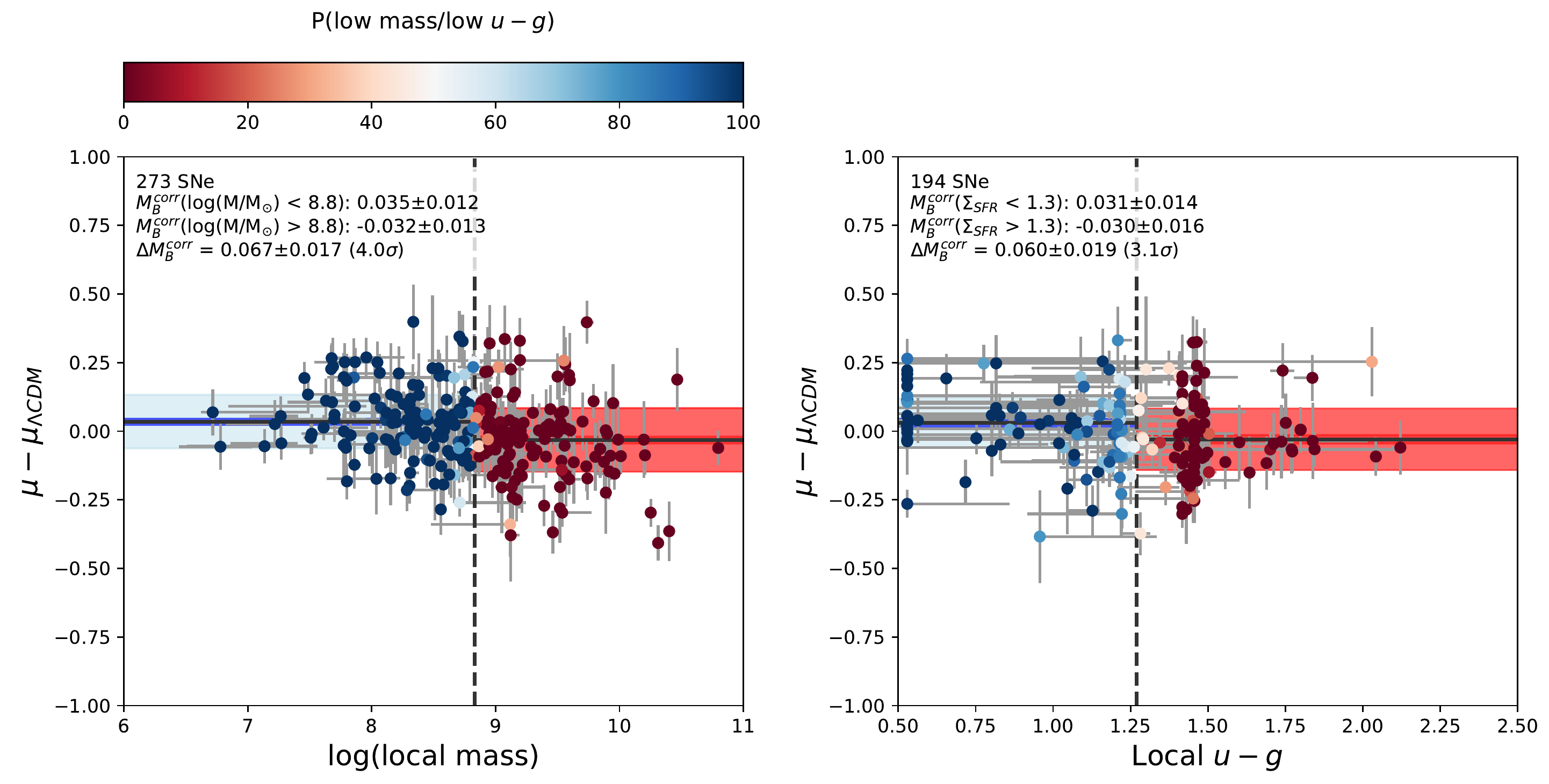}
\caption{The dependence of SN luminosities on the mass
  and $u-g$ color within 1.5 kpc of the SN location.  Colors indicate the
  probability that a SN is in a low-mass host galaxy (left) or a galaxy with
  blue rest-frame color (right).  We see $\gtrsim 2 \sigma$
  correlations with both quantities.  The gap in rest-frame $u-g$ colors
  at $\sim$1.3 mag is likely due to a gap in the colors of the PEGASE.2
  SED templates, the green valley between star-forming and passive host galaxies.}
\label{fig:localstep}
\end{figure*}

\subsection{Measuring the Mass and Color in Random Apertures}
\label{sec:randommeas}

We also consider whether the global step is driven by the local
step and if so, how ``local'' the local measurement needs to be
\citep{Rigault15,Roman18}.
To address this question, we place 150 random apertures
of diameter 3 kpc in each galaxy and measure the local mass and $u - g$ color within
those apertures.  We use the SED template-fitting approach
discussed above to fit the photometry in each aperture individually.
We use these random measurements
to ask whether the region near the
SN is better correlated with SN luminosity than the
regions far from the SN.

We again use the galaxy size- and orientation-weighted SN
separation parameter $R$ to choose where to
place the apertures.  We first use SExtractor to
measure the ellipse that best approximates the shape of a given host
galaxy.  Each region with a given $R$ parameter
lies at the same elliptical radius about the host center.  Regions
with $R = 0$ are at the host center, while regions with $R = 3$ are
approximately at the isophotal limit of the galaxy (shown in
Figure \ref{fig:img}).  Regions with $R = 5$
are outside the isophotal limit of the galaxy and lie far enough away from
the host center that identifying the true host galaxy begins to become
ambiguous.
To include as many apertures near the galaxy
center as far from it, we place random apertures so that 25 have $0 < R < 1$,
25 have $1 < R < 2$, and so on out to $R = 5$, which is the \citet{Sullivan06}
criteria for matching a SN to its likely host galaxy.

We use these random measurements to explore how the local mass and color steps change if host
properties are inferred from regions far from the SN location.  For random
apertures with a given distance
from the SN location or a given $R$, 
we measure the physical properties associated with each SN from the random location
instead of the SN location.  We use these random measurements to find the maximum likelihood
mass and color steps, and compare to the mass and $u-g$ steps using the
properties of the host galaxy at the SN location.  For each set of random
measurements, we choose the median of those measurements for the step
location.  This prevents a situation where the vast majority of the
sample is on one side of the step location, which can occur as apertures move preferentially
towards or away from the host galaxy center.

The spacing of these random apertures
will be less than the seeing of the images in most cases, meaning that
many random measurements will be partially correlated.  However,
we can avoid statistical complications by using just one random measurement per SN at a given time
and avoiding regions within 3 kpc of the true SN location.

\section{Results}
\label{sec:local}

\begin{deluxetable*}{lrrrrr}
\tabletypesize{\scriptsize}
\tablewidth{1.7\columnwidth}
\tablecaption{Mass and Color Step Measurements for Targeted and Non-Targeted Surveys}
\tablehead{&\multicolumn{2}{c}{$\Delta_M$}&&\multicolumn{2}{c}{$\Delta_{u-g}$}\\*[2pt]
\cline{2-3} \cline{5-6} \\*[2pt]
&No Global Mass Corr.&Global Mass Corr.\tablenotemark{a}&&No Global Mass Corr.&Global Mass Corr.\tablenotemark{a}}
\startdata
Local Step&$0.067\pm0.017$&$0.056\pm0.017$&&$0.060\pm0.019$&$0.034\pm0.020$\\
$-$ Targeted SNe&$0.026\pm0.027$&$0.012\pm0.027$&&$-0.001\pm0.030$&$-0.018\pm0.030$\\
$-$ No Targeted SNe&$0.091\pm0.024$&$0.083\pm0.024$&&$0.084\pm0.028$&$0.055\pm0.028$\\
\\
Global Step&$0.058\pm0.018$&$0.001\pm0.018$&&$0.061\pm0.020$&$0.036\pm0.020$\\
$-$ Targeted SNe&$0.061\pm0.034$&$0.003\pm0.035$&&$-0.019\pm0.031$&$-0.032\pm0.029$\\
$-$ No Targeted SNe&$0.049\pm0.024$&$-0.009\pm0.025$&&$0.086\pm0.028$&$0.058\pm0.029$\\
\enddata
\tablenotetext{a}{The size of each step after applying the maximum likelihood global mass correction of \gmstep\ mag.}
\label{table:masscorr}
\end{deluxetable*}

Using the methods described above, we
measure a local mass step of \lmstep\ mag
(\lmstepsig\ significance) and a local color step of \ugstep\ mag
(\ugstepsig).  These steps are shown in Figure \ref{fig:localstep}.
If we use global properties instead of local
to measure the size of the step, we find the global mass step to be \gmstep\ mag and the global
color step to be \gugstep\ mag.
The local mass step is slightly larger than the global mass step, while the
local $u-g$ step is approximately equal to the corresponding global
step.

Table \ref{table:masscorr} summarizes
each global and local step measured from these data,
both before and after correcting for the maximum likelihood
global mass step of \gmstep\ mag.
Most significantly, we find a local mass step of $0.056 \pm 0.017$
mag \textit{after} correcting for the global mass
($3.3 \sigma$).  If we instead correct for the local mass
step before measuring the global step, we find a global
mass step of $0.055 \pm 0.018$ ($3.1 \sigma$).
Table \ref{table:masscorr} also divides the
sample into SNe from surveys that target a pre-selected
set of galaxies and those that do not (\S\ref{sec:targeted} below).

Estimating the statistical significance of the difference between
the global and local steps is complicated by the fact that global and local
measurements are partially correlated.  65\% of the SNe in this
sample are either globally and locally high-mass or globally and
locally low-mass (77\% for local color).
To estimate the 1$\sigma$ uncertainty on the difference between
the global and local step \textit{with correlated measurements}, we simulate 1,000 SN
samples using our real local and global measurements but with
Hubble residuals drawn from a Gaussian centered on 0 and with dispersion equal
to the real dispersion of our sample.  We find that 68\%
of the Monte Carlo samples have a local/global difference $<$ 0.017 mag
for the mass measurements,
and $<$0.032 mag for the color measurements.
These correspond to the 1$\sigma$ uncertainties on the local/global difference,
and are slightly smaller than the uncertainties that would be obtained just
by adding the local and global mass uncertainties in quadrature.
With this approach, we find that sizes of the global and local measurements for both mass and color
are consistent at the 1$\sigma$ level.
Therefore, the data do not indicate that the
local steps are intrinsically more significant than the global steps.

Measuring the local mass step from just the 195 SNe\,Ia with SDSS $u$ data gives a
local mass step of $0.054 \pm 0.019$, 0.013 mag less than the step
measured from the full sample (not statistically significant).
The gap in the dust-corrected, rest-frame
$u-g$ colors (Figure \ref{fig:localstep}) is likely due
to a gap in the colors of the PEGASE.2
SED templates, likely corresponding to the green valley between
passive and star-forming galaxies.
The local and global measurements used in this work
are available online\footnote{\url{http://pha.jhu.edu/~djones/localcorr.html}}.

\begin{figure*}
\includegraphics[width=7in]{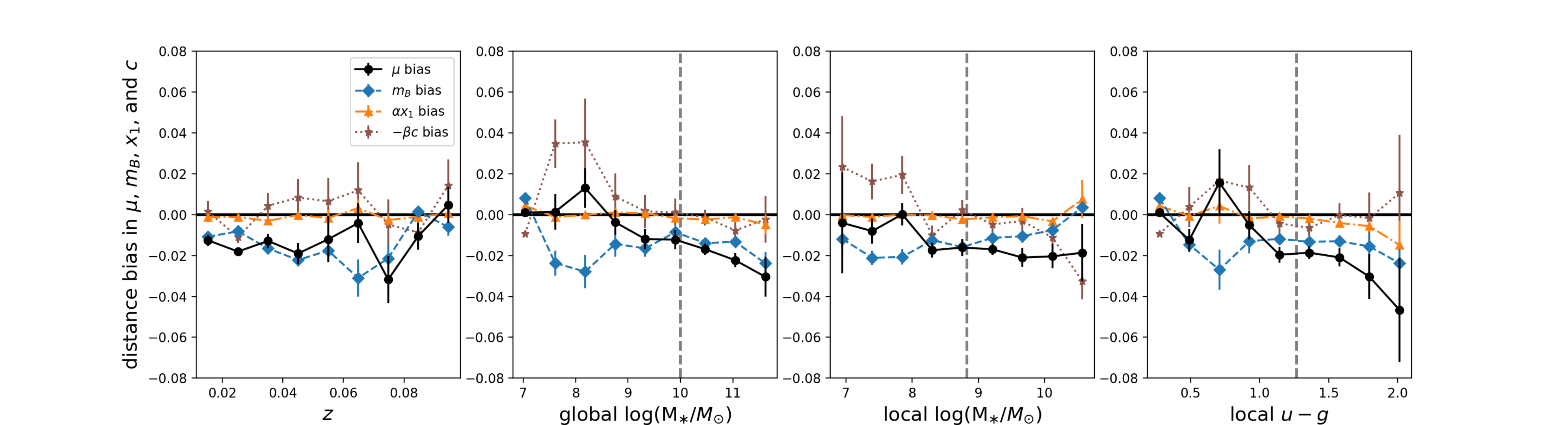}
\caption{The effect of bias corrections on the
  measured host galaxy steps.  In particular, the size of the
  local color step is 27\% larger if the necessary bias corrections
  are neglected, because SN shape and color are functions of host galaxy
  $u - g$.}
\label{fig:biascor}
\end{figure*}

We note that the results are somewhat affected by our method of correcting the sample for biases
  in $x_1$ and $c$.  The BBC method is a relatively new technique that was applied
  in the Pantheon cosmological analysis \citep{Scolnic18}.
  The biases in $x_1$ and $c$ caused by sample selection
  are a clear observational bias that can easily be realized in simulated SN\,Ia samples and found in
  real data \citep{Scolnic14b,Scolnic16}.  The BBC method removes these biases and in
  doing so reduces the local u-g step by 27\% due to the strong dependence of SN\,Ia shape and color on galaxy
  properties.  We show this dependence for the sample presented here
  in Figure \ref{fig:biascor}.  The apparent size of the
  local mass step decreases by just 4\%.

\subsection{Targeted Versus Untargeted Surveys}
\label{sec:targeted}

\begin{figure*}
\includegraphics[width=7in]{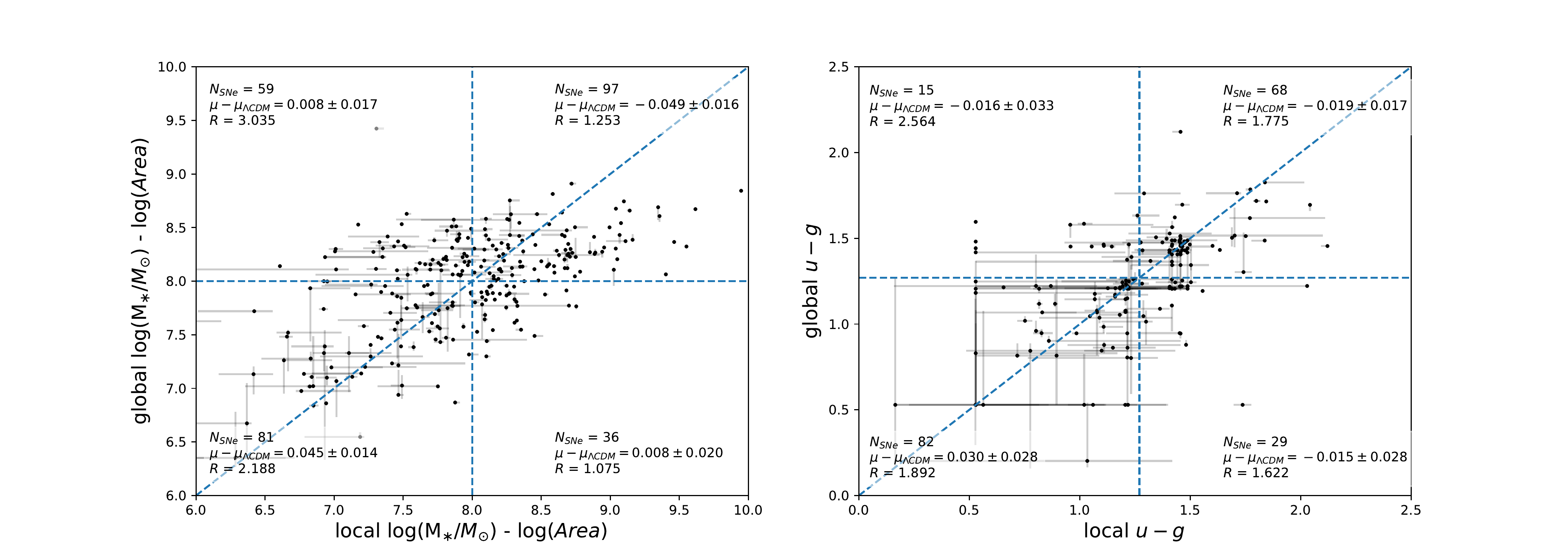}
\caption{Local versus global information for each SN\,Ia in the sample,
  where dashed lines indicate the median mass and
  color along with the points where local measurements
  equal global measurements.
  The median Hubble residual for each quadrant is labeled, and
  shows significant departures from 0 only where local and global agree.
  Instead of the local and global masses used in this work,
  we show the mass density, log(M$_{\ast}$/M$_{\odot}$) - log(Area), where
  the units of area are kpc and the global mass density is averaged over the total area within
  the isophotal radius ($R = 3$) of each galaxy.}
\label{fig:localvglobal}
\end{figure*}

\citet{Roman18} measured a step from $z < 0.1$ SNe\,Ia
that was $0.038 \pm 0.034$ mag smaller
than the step they measured from the full
sample, though the difference was not statistically significant.
If confirmed, this difference could either be due to
a redshift evolution of the local step or differences in
low-$z$ versus high-$z$ survey methodology.  Specifically, much of the low-$z$ data are from surveys
that target a pre-selected set of (usually NGC) galaxies.
None of the high-$z$ surveys
target pre-selected galaxies.
Targeted surveys also collect SNe that are more like the sample of SNe\,Ia within
$\sim$40 Mpc that are calibrated by Cepheids and used as a
rung on the distance ladder for measuring H$_0$.  On the other hand, all $z > 0.1$ data
used for measuring the dark energy equation of state
come from surveys that do not target specific galaxies.
It may also be relevant that the CfA and CSP low-$z$ SNe were observed
on the Johnson filter system, while Foundation and the
$z > 0.1$ data were primarily observed on the Sloan filter system.
Because SNe observed on the Sloan filter system
have $g$ as the bluest band, there could be 
differences if host galaxy biases affect SN\,Ia luminosity in a 
wavelength-dependent manner.

Because because Foundation data come predominantly
from untargeted surveys (Gaia, ASAS-SN, PSST),
our data can be used to determine whether SNe from targeted surveys have
a different local or global step than SNe from untargeted surveys.
Foundation includes some
data from targeted surveys only because untargeted surveys would
likely discover these SNe if the targeted surveys did not exist \citep{Foley18}.
We therefore treat Foundation as an untargeted survey in this analysis.

In Table \ref{table:masscorr} we compare the local and global steps
measured from $z < 0.1$ SNe in targeted surveys (CfA and CSP) and $z < 0.1$ SNe from
surveys that are not targeted (Foundation, PS1, and SDSS).
After global mass correction, We see a 2.0$\sigma$ increase in
the local mass step, a 1.8$\sigma$ increase in the local color step,
and a 2.1$\sigma$ increase in the global color step
when untargeted surveys are used instead of targeted surveys.  Only the difference
in the global mass step is statistically insignificant.
These differences are not highly significant but
could indicate that the correlation of SN distance with
host galaxy properties is sensitive to survey selection effects.
In the Appendix, we examine the differences in intrinsic dispersion
  on either side of the mass and color divide, finding 1-2$\sigma$ evidence
that SNe in locally low-mass or locally blue regions may have lower dispersions.

One might expect the difference in significance to be affected by the fact that
  SN\,Ia in targeted surveys could be biased towards
  regions with higher stellar mass.  Though this is the case for the global
  mass, it is not the case for the local mass as many SN\,Ia in the targeted
  sample are far from the centers of their host galaxies.
  Of the SN\,Ia in targeted surveys used in this study, 83\% (91 of 110)
  have global masses $> 10$ dex, the global mass step divide.  However,
  51\% (56 of 110) occurred in locally massive regions
  locally massive (local mass $> 8.83$ dex).

We also check the significance of a local step vs a global step
using the Foundation sample alone.
Our sample includes 127 Foundation SNe with $grizy$ data that can
be used to measure the local mass step and 80 Foundation
SNe with SDSS $u$ observations that can be used to
measure the local color step.  We find a local mass step
of \flmstep\ mag (\flmstepsig) and a local color step of \fugstep\ mag (\fugstepsig).
We find a global mass step of \fgmstep\ mag and a global color step of \fgugstep\ mag,
both consistent with the local steps.
We find a 2.1$\sigma$ difference between the Foundation and non-Foundation
local mass step and 2.9$\sigma$ difference between the Foundation and non-Foundation
local color step.

\begin{deluxetable*}{lrrr}
\tabletypesize{\scriptsize}
\tablewidth{1.7\columnwidth}
\tablecaption{Combining Local and Global Steps}
\tablehead{&Local Step&Global Step&Combined Step}
\startdata
local mass, global mass&0.059$\pm$0.019&0.046$\pm$0.021&0.105$\pm$0.025\\
local $u-g$, global $u-g$&0.046$\pm$0.028&0.025$\pm$0.025&0.070$\pm$0.030\\
local mass, global $u-g$&0.046$\pm$0.025&0.046$\pm$0.023&0.092$\pm$0.031\\
local $u-g$, global mass&0.030$\pm$0.033&0.039$\pm$0.032&0.069$\pm$0.028\\

\enddata
\label{table:twostep}
\end{deluxetable*}

\subsection{Varying Nuisance Parameters}
\label{sec:nuisance}

The correlation of SN shape and color may also change
  as a function of host galaxy properties; $\beta$, in particular,
  could be subject to change due to the change of dust properties
  as a function of host mass or color.  For this reason, we tested the effect of adding
  separate $\alpha$ and $\beta$ parameters to the likelihood model for each
  side of the mass or color step.

\begin{figure}
\includegraphics[width=3.5in]{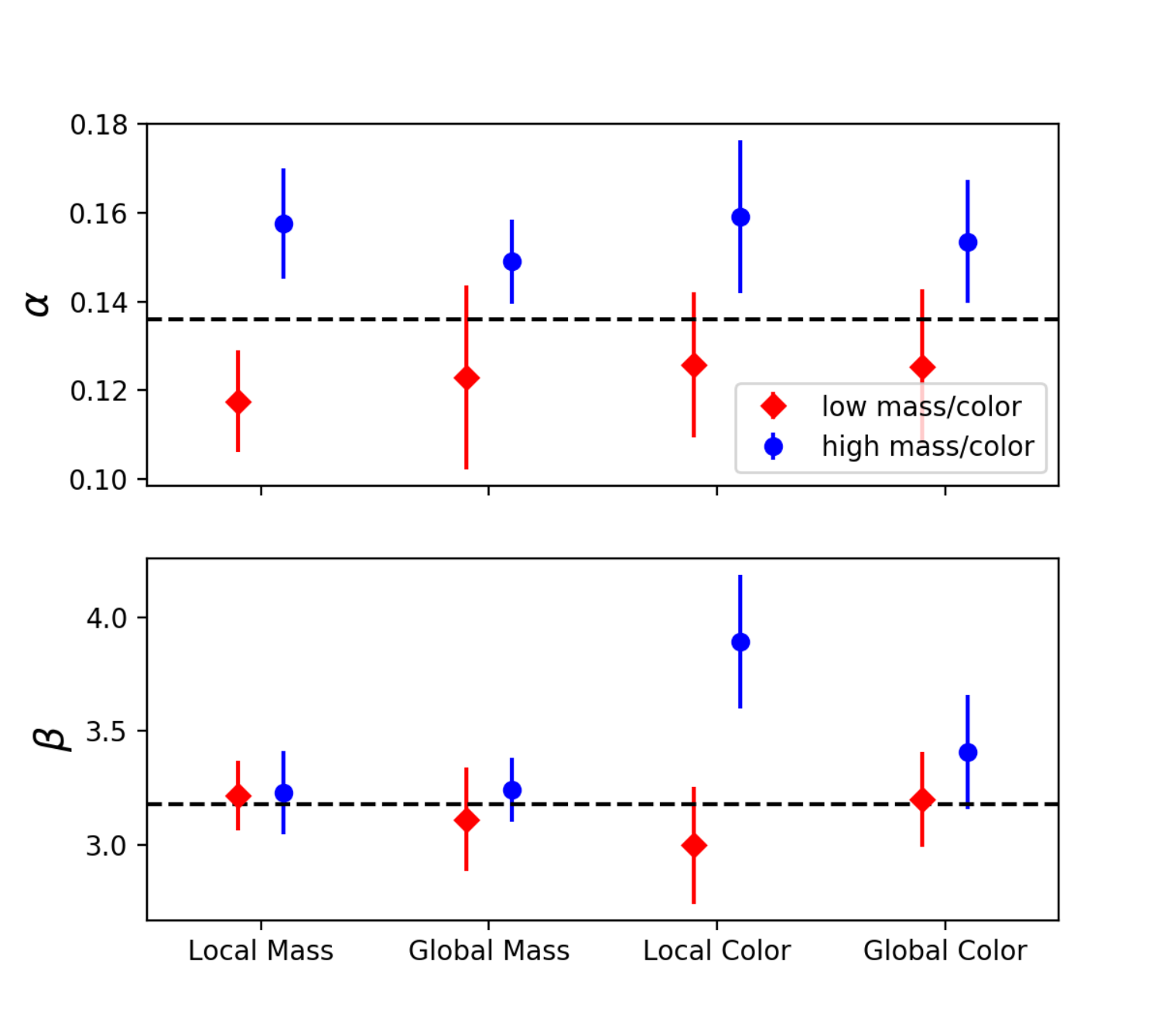}
\caption{The dependence of nuisance parameters $\alpha$ and
  $\beta$ on host mass and color.  Interestingly, $\alpha$ is
  measured to be higher in locally/globally red or high-mass hosts.
  $\beta$ is higher (nearer to the Milky Way value) for
  SNe that occurred in redder regions of their host galaxies,
  likely due to dust effects.}
\label{fig:alphabeta}
\end{figure}

  The results of $\alpha$ and $\beta$ variation are shown in Figure \ref{fig:alphabeta}.
  We find that $\alpha$ is universally higher in locally red regions and
  locally massive regions ($\sim 2 \sigma$ significance).  We find a significant difference
  in $\beta$ only for SNe in locally red regions of their host galaxies, which most likely
  implies that the effect is driven by dust obscuring the SN.

  When $\alpha$ and $\beta$ are allowed to vary, the local mass step
  increases by 0.005 mag, while the local color step increases by 0.038
  mag.  Both increases are only marginally significant,
  but similarly to \citet{Rigault18}, we find that
  allowing $\alpha$ and $\beta$ to be fit simultaneously with the
  local or global step tends to increase the size of the step and reduce
  the dispersion.  We find a dispersion of just 0.047 mag for SNe in locally
  \textit{red} regions but with a high uncertainty, such that the difference
  between locally red versus blue regions is not statistically significant.
  Previous results, e.g. \citet{Rigault15,Kelly15} have found lower dispersion
  in locally \textit{blue} regions, but did not allow $\alpha$ and $\beta$ to vary (our sample also gives this result
  at 1$\sigma$ significance).

\subsection{Simultaneously Fitting a Global and Local Step}

Table \ref{table:masscorr} shows that after global mass correction,
only the local mass step remains significant
at $>$3$\sigma$ ($0.056 \pm 0.017$ mag).  Previous studies (e.g. \citealp{Roman18}), have
seen a similar effect, which they interpret as evidence that local regions encode information
about the SN progenitor that is not captured by a global correction.

In Figure \ref{fig:localvglobal} we show the relationship between the
local and global measurements in this work to understand which SNe
are being corrected by the global versus the local steps.
We show the global and local mass densities instead of the global and local mass
used elsewhere in this analysis, in order for the local and global units to be the same
in this figure.
In particular,
there are a number of SNe far from the centers of their host galaxies that
have high global mass densities but low local mass densities.  We label the
weighted average of the Hubble residuals in each quadrant. If the local step were driving the global step,
we would expect to see a change in Hubble residual only along the x-axis (the local measurement
axis).  Similarly, if the global measurement were driving the local correction,
we would expect the average Hubble residual to change only along the y-axis.  Instead, we
see $\sim$4$\sigma$ evidence (mass) and $\sim$2.6$\sigma$ evidence (color)
for a step when considering \textit{only}
the two quadrants where local and global agree.

In the previous sections, we have measured only a single step
at a time.  Beginning with the standard likelihood approach presented in \S\ref{sec:analysis}, we
now expand the method
to simultaneously measure a combined local and global step for mass and color.  The results
are summarized in Table \ref{table:twostep}.  By measuring global and local mass steps together,
we find a 3.1$\sigma$ local mass step
and a 2.2$\sigma$ global mass step.  The intrinsic dispersion about the Hubble diagram (the
dispersion after photometric uncertainties are taken into account) is
3-5\% lower than the dispersion after correcting for a single step.  The combined
local and global $u-g$ step is less significant than the mass step.

The evidence for a combined local/global mass step is
marginally significant, with a Bayesian Information Criterion
that is slightly lower ($\Delta$BIC = 3.4)
when an extra step is included in the likelihood model.
Making either a global step or local step alone leaves an additional step
with $\gtrsim 3 \sigma$ significance.  Therefore, the possibility that local
and global may reinforce each other is intriguing.

\subsection{Random Apertures}

\begin{deluxetable*}{lrrrrr}
  \tabletypesize{\scriptsize}
  \tablewidth{1.7\columnwidth}
\tablecaption{Comparing Local to Random Measurements}
\tablehead{&\multicolumn{2}{c}{$\Delta_M$}&&\multicolumn{2}{c}{$\Delta_{u-g}$}\\*[2pt]
\cline{2-3} \cline{5-6} \\*[2pt]
&No Global Mass Corr.&Global Mass Corr.\tablenotemark{a}&&No Global Mass Corr.&Global Mass Corr.\tablenotemark{a}}
\rowcolor{gray!25}Local Step&$0.067\pm0.017$&$0.056\pm0.017$&&$0.060\pm0.019$&$0.038\pm0.019$\\
\rowcolor{gray!25}Random Step&$0.047\pm0.018$&$0.029\pm0.017$&&$0.040\pm0.020$&$0.027\pm0.020$\\
5 kpc from SNe&$0.025\pm0.019$&$0.000\pm0.018$&&$0.031\pm0.021$&$0.012\pm0.021$\\
10 kpc from SNe&$0.032\pm0.021$&$0.015\pm0.020$&&$0.051\pm0.021$&$0.036\pm0.021$\\
$R < 1$&$0.021\pm0.019$&$0.002\pm0.019$&&$0.039\pm0.023$&$0.011\pm0.023$\\
$1 < R < 2$&$0.041\pm0.019$&$0.018\pm0.019$&&$0.046\pm0.021$&$0.031\pm0.021$\\
$2 <  R < 3$&$0.044\pm0.018$&$0.021\pm0.019$&&$0.053\pm0.020$&$0.037\pm0.020$\\
\tablenotetext{a}{The size of each step after applying the maximum likelihood global mass correction of \gmstep\ mag.}
\tablenotetext{b}{Regions $>$5 kpc from SN are randomly sampled.  One
  random region is chosen per SN, the step is measured, and this
  process is repeated 100 times.  The steps listed
  here are the mean of 100 samples.}
\tablecomments{$R$ is the distance from the center of the galaxy
  in units of the normalized elliptical radius of the galaxy \citep{Sullivan06}.  The last 5
  rows exclude regions within 3 kpc of the SN location.  Also in the last 5 rows,
  the step location is taken to be the median of every sample to avoid a situation in which
  90\% or more of the sample is considered ``high-mass'' or ``low-mass''.}
\label{table:random}
\end{deluxetable*}

\begin{figure*}
\includegraphics[width=7in]{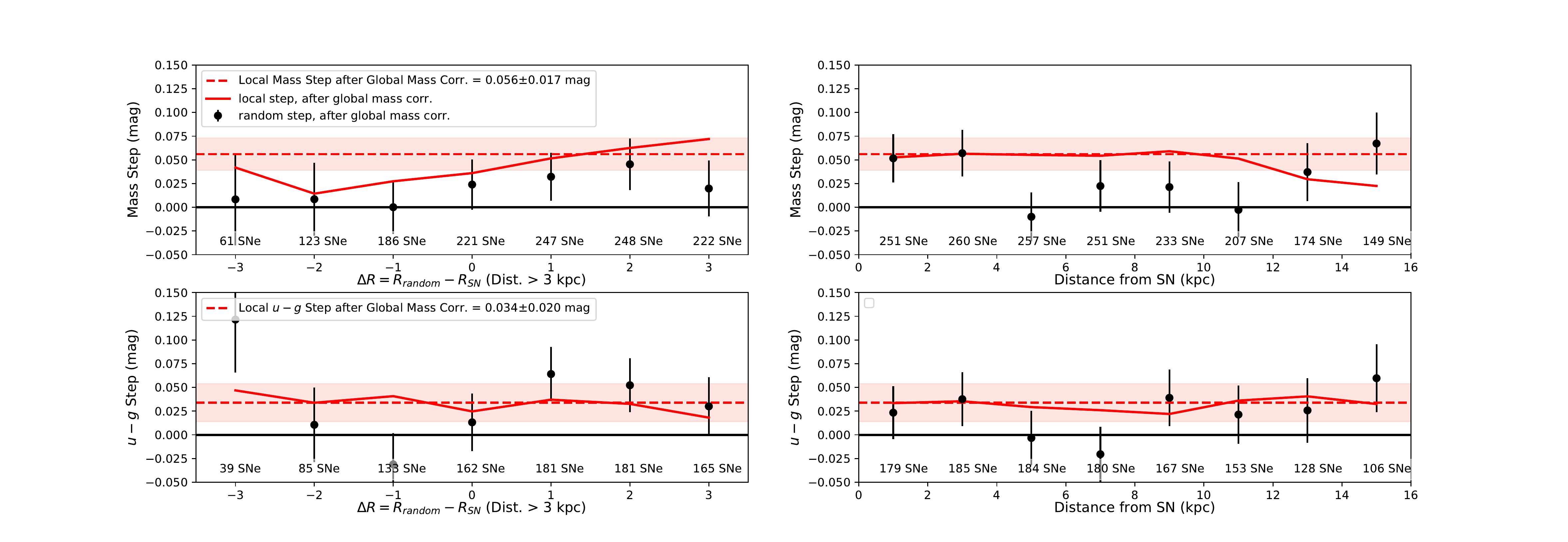}
\caption{After global mass correction, the ``false local step'' (black): the correlation of
  SN distance measurements with the masses and $u-g$ colors of
  different regions in the host galaxy.  $\Delta R$ is the difference in $R$ between
  the SN location and the random location after excluding all regions within 3 kpc of the
  true SN location.  The local step after global mass correction
  and its uncertainty are indicated by the shaded region.
  For each false local step, the true local step at the SN location (red line)
  is plotted using the \textit{same set of SNe used to measure the random step}.}
\label{fig:randomlocal}
\end{figure*}

Having seen evidence for a local mass step after global
mass correction, the question remains how ``local''
the local measurement would need to be to correct SN distances.
To answer this question, we use the measurements of mass and
color within random apertures discussed in \S\ref{sec:randommeas}.
We summarize the results of these random tests in Table \ref{table:random}.
By inferring local properties from random regions $>$5 kpc from the
SN location after first correcting for the global mass step, we
measure a ``false local'' mass step of $0.029 \pm 0.017$ mag.  This step is smaller
than the true local step by
$0.027 \pm 0.017$ mag, a difference with 1.6$\sigma$ significance.  As discussed at
the beginning of \S\ref{sec:local}, these
uncertainties incorporate
the correlation between the local and random measurements.
We measure a $u-g$ step of $0.027 \pm 0.015$ mag,
$0.011 \pm 0.027$ mag smaller than the local step.
We therefore see only marginal evidence that measurements of
host galaxy properties within 5 kpc of the SN location are
important for SN distance corrections.
In the Appendix, we examine the differences in intrinsic dispersion
  between local and random regions, finding no significant difference in dispersion
  when local mass and color are inferred from random locations instead of locations
  near the SN.

The last five rows of Table \ref{table:random} show false local steps using a set of representative $R$
parameters and distances from the SN.  \textit{All} of these measurements yield steps smaller than the
local step, typically by $\sim 2 \sigma$ significance for mass and $\sim 1 \sigma$ for color.
The $R$ measurements in Table \ref{table:random} do not include regions within 3 kpc
of the SN, so that no measurements include the true local
fluxes at the SN location.  We also restrict distance measurements to $R < 5$.

Figure \ref{fig:randomlocal} expands the results in Table \ref{table:random} to
show change in the local mass and color steps
as a function of both $\Delta R$, the difference in $R$ between the SN and the aperture (left),
and of the aperture's physical distance from the SN (right).
Negative $\Delta R$ indicates that physical properties are inferred from
regions closer to the galactic center than the SN location, while
positive $\Delta R$ means that the physical properties are inferred from
regions farther from the galactic center than the SN location.

As distances from the SNe increase, the sampling
of random apertures becomes slightly more sparse and therefore the mass and color steps
are not always computed using the full SN sample.  There is a similar effect in play
for different values of $\Delta R$; for a SN at the center
of its host galaxy, having a random aperture with $\Delta R < 0$ is impossible.  Similarly,
a SN near the edge of its host could not have a large $\Delta R$.  Small hosts in particular will have
a restricted range of $\Delta R$ and physical distances $>$10 kpc from the SN location
may be outside the $R = 5$ ellipse.  Therefore, there are significant biases in the global
host demographics for different $\Delta R$ parameters and distances.
For this reason, in Figure \ref{fig:randomlocal} we always compare the false local steps to the
true local steps measured using the exact same set of SNe.

There are hints that the SN distance measurement becomes less
correlated with the localized host galaxy mass at $\gtrsim$5 kpc from
the SN.
We also find that a number of mass step measurements are
smaller than the local step by $\gtrsim$0.03 mag ($\sim 2 \sigma$).
The statistical significance of these differences is limited and
different $\Delta R$ steps are not completely statistically independent.
However, the observed differences between random and local are consistent with
the observed $0.056 \pm 0.017$ mag local mass step after global mass
correction.  However, we see no statistically significant
difference between the local and random color step.

We find a 0.011 mag decrease in the random color step compared to the local
color step, which is consistent with \citet{Roman18}.  \citet{Roman18} find
a decrease in the size of a local color step of 0.022 mag when changing from
their nominal local radius of 3 kpc to a radius of 16 kpc, approximately
the maximum distance from the SN location considered here.  Because we use
only a low-$z$ sample to examine local regions, our uncertainties are larger
than those of \citet{Roman18}, and a difference of 0.022 is comparable
to the 1$\sigma$ local color uncertainties.
However, the 1$\sigma$ uncertainties on this test constrain the effect of
a non-local measurement to $\lesssim$0.04 mag.

\subsection{Local Specific Star Formation Rate}
\label{sec:ssfr}

\begin{figure}
\includegraphics[width=3.5in]{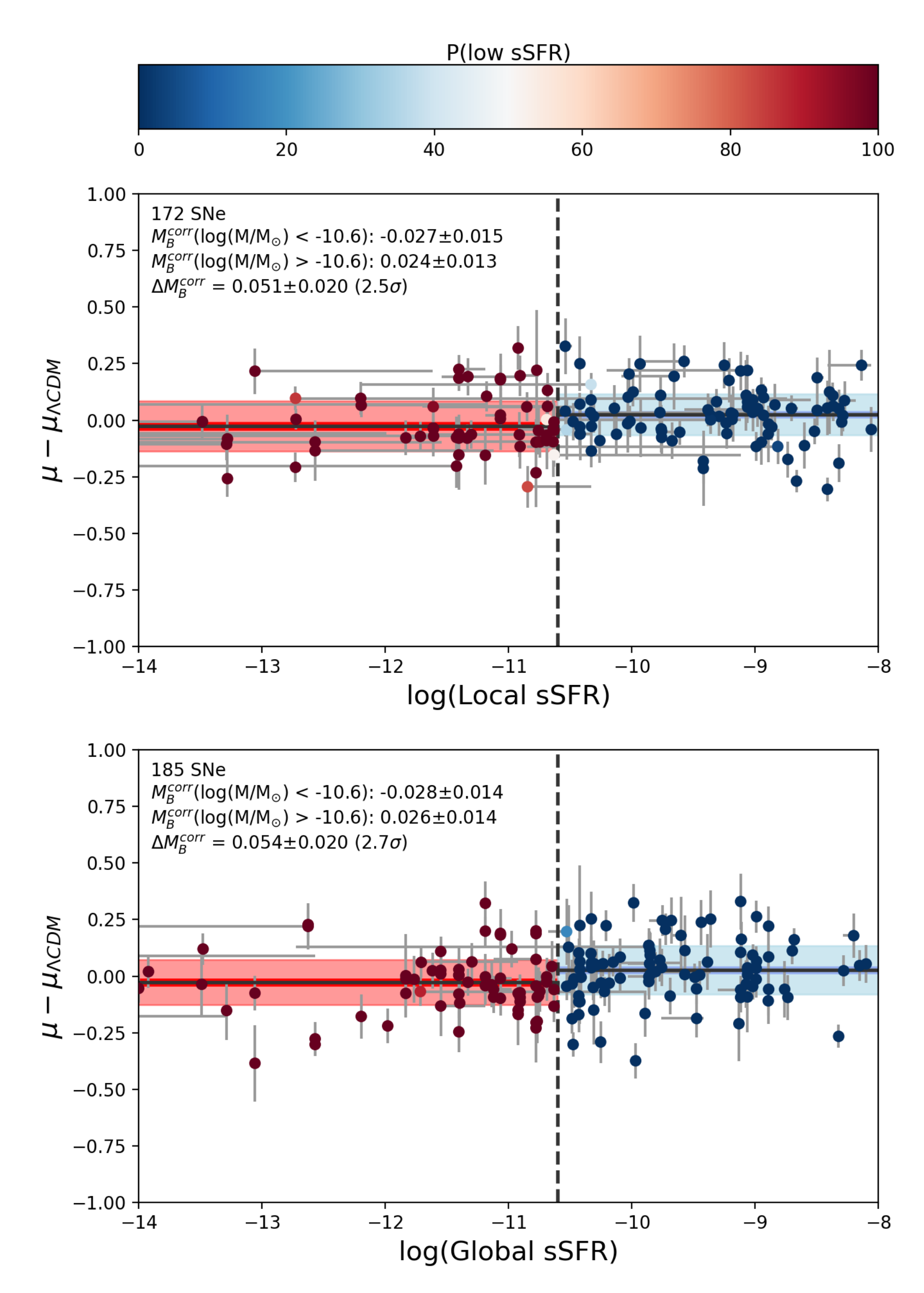}
\caption{Correlation of Hubble residuals with local and
  global sSFR.  We observe steps at $2.5-2.7 \sigma$ significance,
  with no significant difference between the local and global steps.
  After global mass correction, we find a local step of $0.035 \pm 0.021$ and a
global step of $0.029 \pm 0.020$.}
\label{fig:lssfr}
\end{figure}

Recent work from \citet{Rigault18} has suggested
that the local specific star formation rate (LsSFR) has
a strong correlation with SN\,Ia residuals in SNFactory
data.  Though we lack the local H$\alpha$ measurements
used by \citet{Rigault18}, $ugriz$ photometry should
enable us to investigate whether such a correlation
is present in our data, though we caution that H$\alpha$
may be a more robust diagnostic of sSFR.

As Z-PEG has difficulty measuring sSFR in 
passive galaxies, we use LePHARE \citep{Arnouts11} with \citet{Bruzual03}
spectral templates as an alternative
SED-fitting method to infer LsSFR from our sample.
We also use LePHARE for consistency checks on our
data in \S\ref{sec:consistencychecks} below.
The SED-fitting parameters from LePHARE are broadly
consistent with those from Z-PEG; the median $u-g$
color from LePHARE is within 0.04 mag
of the Z-PEG color and the median host galaxy mass is
just 0.09 dex lower.

Using the median LsSFR of our sample, -10.6 dex,
as the divide between SN\,Ia in low- and high-sSFR
regions, we measure a step size of $0.051 \pm 0.020$ mag (2.5$\sigma$).  The
global sSFR step size is nearly identical, $0.054 \pm 0.020$ mag (2.7$\sigma$).
The significance of both steps becomes $< 2 \sigma$ after
global mass step correction (the local step becomes $0.035 \pm 0.021$ mag and the global
step becomes $0.029 \pm 0.020$ mag).  \citet{Rigault18}
find a step of $0.125 \pm 0.023$ mag ($0.163 \pm 0.029$ mag after allowing
$\alpha$ and $\beta$ to be fit simultaneously with the step).  Their result is
statistically inconsistent with ours at the 2.4$\sigma$ level,
though removing the BBC corrections would reduce that discrepancy to 2.1$\sigma$.
If this discrepancy is not due to statistical fluctuation or unforeseen
systematic effects due to differences in sample selection, calibration, or sSFR
measurement methods, it may be additional evidence that targeted versus
untargeted surveys affect the measured step sizes.
Interestingly, the sSFR step sizes we measure do not significantly
change if we use only SNe from targeted or untargeted samples.

We show the local and global LsSFR steps in Figure \ref{fig:lssfr}
and include the LsSFR measurements from LePHARE in our
online data.

\subsection{Consistency Checks}
\label{sec:consistencychecks}

  In this section we present several consistency checks to
  validate the SED fitting procedures in this work.  The \texttt{Z-PEG}
  SED fitting method is significantly different than that of
  \citet{Roman18}, for example, who also base their results on the
  \texttt{PEGASE.2} templates but warp those templates to match
  the observed photometry of galaxies in the SuperNova Legacy
  Survey fields.

  However, 54 SN\,Ia in this sample are also included in
  \citet{Roman18}.  We compare our
  rest-frame U-V colors and observed $u-g$ colors to those measured by \citet{Roman18}
  using their online data.  Though we measure $u-g$ colors within a 1.5 kpc radius
  while \citet{Roman18} use a 3 kpc radius, we observe a median color of 1.52 mag, just
  0.06 mag bluer than that of \citet{Roman18}.  Though we use rest-frame $u-g$ colors in this work,
  after fitting with \texttt{Z-PEG}, we verified that
  the rest-frame $U-V$ colors were consistent with \citet{Roman18}: we find a median rest-frame
  $U-V$ color of 0.83 mag compared to 0.77 mag for \citet{Roman18}.
  If we measure a $U-V$ step instead of a $u-g$ step, we find that the step
  size increases by just 5 mmag.
  
  Z-PEG returns a set of ``pseudo-observed'' model magnitudes, which have
  been reddened and redshifted to match the observed data.  These model
  magnitudes should be close to the observed data if our SED fitting
  procedure is reliable.  For 141 of the 194 SN\,Ia with $u$ observations,
  the pseudo-observed magnitudes are within the 2$\sigma$ uncertainties
  on the local $u-g$ color observations.  In reality, there is some additional
  uncertainty on the model which would increase the statistical agreement between model
  and data.  If we restrict our sample to just these 141 SN\,Ia, we measure
  a local color step of $0.044 \pm 0.022$ mag, 
  but consistent with the $0.060 \pm 0.019$ mag step measured from the full sample.

  For the local mass step, a simple consistency check may be performed using
  the relationship between $gi$ photometry and host galaxy stellar mass given
  by \citet{Taylor11}:

  \begin{equation}
\mathrm{log}(M_{\ast}/M_{\odot}) = 1.15 + 0.70(g - i) - 0.4M_i,
    \end{equation}

  \noindent where the absolute $i$ magnitude M$_i$ is estimated using $\Omega_M = 0.3$,
  $\Omega_{\Lambda} = 0.7$ and H$_0$ = 70 km s$^{-1}$ Mpc$^{-1}$.  The uncertainty of the
  relation is 0.1 dex, which we add in quadrature to the propagated photometric
  uncertainty.
  
  Although this equation does not k-correct the photometry, it is still a
  reasonable approximation for these low-redshift data.  Using this approximation
  instead of \texttt{Z-PEG}, we measure a local mass step of $0.077 \pm 0.017$ mag,
  consistent with the measurement of $0.067 \pm 0.017$ mag measured from the full sample.

  Lastly, we use the \texttt{LePHARE} SED-fitting software \citep{Arnouts11} with \citet{Bruzual03}
  templates (the version used for the COSMOS mass function; \citealp{Ilbert09})
  to independently check the mass and color measurements from Z-PEG.  We find a median
  color just 0.04 mag redder than the \texttt{Z-PEG} measurements and a median host mass
  0.09 dex smaller than the \texttt{Z-PEG} measurements.  LePHARE yields less model-dependent
  colors than \texttt{Z-PEG}, as it uses the SED templates for $k$-corrections but interpolates
  using those $k$-corrections from the observed magnitudes themselves.  We measure local mass
  and color steps that are consistent with, though slightly smaller than, the \texttt{Z-PEG}
  measurements: with \texttt{LePHARE}, we measure a nearly identical local mass step
  of $0.066 \pm 0.017$ mag and a local color step of $0.047 \pm 0.019$ mag.

\begin{deluxetable*}{lrrrr}
\tabletypesize{\scriptsize}
\tablewidth{1.7\columnwidth}
\tablecaption{Predicted Change in H$_0$ due to Mass and Color Steps}
\tablehead{&Step Significance\tablenotemark{a}&\% in Cepheid Calibrators&\% in Hubble Flow&$\Delta$H$_0$ (km s$^{-1}$ Mpc$^{-1}$)}\\*[2pt]
\startdata
local mass $>$ 8.83 dex&$3.2\sigma$&36.8&52.1&-0.28\\
global mass $>$ 10 dex&$0.1\sigma$&47.4&70.0&0.02\\
local $u-g$ $>$ 1.27&$1.7\sigma$&10.5&50.0&-0.44\\
global $u-g$ $>$ 1.27&$1.8\sigma$&26.3&46.5&-0.24\\
local sSFR $<$ -10.6&$1.7\sigma$&21.1&52.0&-0.37\\
global sSFR $<$ -10.6&$1.4\sigma$&31.6&52.7&-0.21\\
\enddata
\label{table:h0}
\tablenotetext{a}{Significance of the step after 0.06 mag correction based on global mass.}
\tablecomments{We show the effect of applying a local step after correcting
  for a 0.06 mag mass step following \citet{Riess16}.  We note that the H$_0$
  correction appears to be stronger in untargeted surveys of SNe\,Ia than it does in targeted surveys
  such as the \citep{Riess16} sample.  Note that the ``global mass''
  correction increases H$_0$, as we measure a slightly smaller mass step of 0.058 mag
  in this work.  However, the steps applied are nearly identical to those listed in
  the ``Global Mass Corr.'' columns of Table \ref{table:masscorr}.}
\end{deluxetable*}

\section{Impact on the Hubble Constant}
\label{sec:cosmoparam}

A leading approach for measuring the Hubble Constant, H$_0$,
calibrates the luminosity of SNe\,Ia in nearby galaxies using Cepheid variables
and compares them to SNe\,Ia in the Hubble flow (typically $z \gtrsim 0.01-0.02$).
A potential bias may enter if there are differences in the mean host
properties of the two SN samples
for some of the host properties considered here.

The determination of H$_0$ in \citet{Riess16} corrects the two
SN samples for the global mass step
using a value of 0.06 mag \citep{Betoule14},  consistent with the $0.058 \pm 0.017$ mag global
step we measure in this work.
After the 0.06 mag global mass step is applied to our sample,
instead of the 0.058 mag global mass step determined in \S\ref{sec:local},
we measure residual, local step sizes 
of $0.055 \pm 0.017$ mag (mass) and $0.033 \pm 0.020$ mag (color).  Of
these, only the local mass step
may be considered significant and may indicate a bias.  Here we
calculate the size of a possible bias in H$_0$.
We also note that for a local step to resolve the
discrepancy between the local measurement and the CMB-inferred
value \citep{Planck15}, the effect would also have to be present in SN\,Ia J-band
luminosity \citep{Dhawan18}.

We use the method developed by \citet{Rigault15} (and also used in \citealp{Jones15b})
to calculate the potential bias to H$_0$ due to a local step.  The bias to the Hubble constant due to a local
mass step is given by:

\begin{equation}
\label{eq:H0}
 \log(H_0^{\mathrm{corr}}) = \log(H_0)
 - \underbrace{
   \frac{1}{5}(\psi^{HF}-\psi^{C})\times \DcHR,}_{\text{local bias correction}}
\end{equation}

\noindent where $\DcHR$ is the size of the local step after removing the global step.
$\psi^{HF}$ and $\psi^{C}$ are the fractions of SNe\,Ia in the Hubble flow
and in galaxies with Cepheid observations, respectively,
that occurred in locally massive regions
of their hosts.  We use the recent measurement of
H$_0 = 73.48\pm1.66$ from \citet{Riess18} as our baseline.
$\psi^{HF}$ is computed using only the SNe in this analysis that are
also included in \citet{Riess16}.
16 of 19 total Cepheid calibrators have PS1 imaging,
as 3 (SN 2001el, SN 2012fr, and SN 2015F) are too far south for PS1.
An additional 2 SNe lack SDSS $u$ imaging (SN 2005cf and SN 2007sr).
For these 5 SNe, we use SkyMapper photometry \citep{Wolf18} instead of PS1 and
SDSS photometry to determine the local masses, global colors, and local colors.

Because the fraction of SNe Ia with local masses above or below the
step is fairly well balanced across the Cepheid calibrator and Hubble flow samples, with a fractional
sample difference of $\sim 0.15$, the effect on H$_0$ is a small
fraction of the step, reducing it by $0.28$  km s$^{-1}$
Mpc$^{-1}$.  This shift is 17\% of the present uncertainty in $H_0$.
A slightly larger sample difference is seen for local $u-g$ colors.
We find that
89.5\% of Cepheid calibrators are in $u - g < 1.6$ galaxies.
In contrast, $\sim$50.0\% of the Hubble flow sample
are in $u - g < 1.6$ galaxies.
However, because the significance of the local color step (after
global mass correction) is just $1.7 \sigma$, no correction is warranted.

For the local mass, global mass, local $u-g$ and global $u-g$ steps,
Table \ref{table:h0} gives the estimated bias to H$_0$ using the
measurements in this work after a global mass correction.  These range from 0.02 to -0.44 km s$^{-1}$
Mpc$^{-1}$.  However, only the local mass step is significant and thus could be considered meaningful.

A caveat to applying even the local mass step correction may
be drawn from the differences in steps suggested in the previous
section for targeted and non-targeted surveys.
Both the Cepheid calibrated and Hubble flow samples used in
\citet{Riess16} came exclusively from targeted surveys in which
all local steps with or without the global mass correction applied are smaller and
not significant with only $\sim 1 \sigma$ confidence.  If the
present hint of a difference in step sizes between these survey
types is established with larger surveys, we would conclude that
no additional correction to $H_0$ would be warranted for these
local steps.  At present a conservative approach would be to
apply half the shift to H$_0$ and consider half the shift as
part of the systematic uncertainty.

An alternative approach to accounting for differences in the
host properties of SN samples could be to ensure both samples are
homogeneous.  For the determination of $H_0$ using Cepheids
to calibrate SNe Ia, it is necessary to select
calibrators from late-type galaxies.  Placing this same
selection criterion on the Hubble flow sample, as done
in \citet{Riess16}, has a negligible impact on the uncertainty
in $H_0$ because the number of SNe Ia in late-type hosts in the
Hubble flow is much larger than the number of calibrators.

\section{Conclusions}
\label{sec:conclusions}

We used up to 273 SNe from the Pantheon and
Foundation samples to determine whether
the physical properties of the regions near the location of SNe\,Ia
are as correlated with SN light curve
parameters and inferred SN distances as global host properties or
random regions within those same host galaxies.
This sample is $\sim$40\% larger than the low-$z$ sample used in recent
measurements of cosmological parameters.
Our measurements of local masses and local, rest-frame $u - g$ colors
for the full sample are available online\footnote{The
  data are available at \url{http://pha.jhu.edu/~djones/localcorr.html}.}.

We see a significant correlation between local stellar mass
and SN distance residuals.  The presence of a $0.056 \pm 0.017$
mag local mass step after global mass correction is
compelling evidence that local effects should be explored
in future analyses.
However, even with the largest
sample of $z < 0.1$ SNe\,Ia to date,
were unable to definitively prove that local information is
better-correlated with SN\,Ia distance measurements
than global or random information.  We found just 1.6$\sigma$ evidence that
SN\,Ia Hubble residuals were better correlated with local information than
with random information inside the same host galaxy.

We find evidence for a correlation
between Hubble residuals of SNe for which local and global measurements
agree.  The difference between the inferred distances of SNe in
both locally high-mass regions and globally high-mass galaxies versus those
in locally/globally low-mass regions is $0.105 \pm 0.025$ mag.
The evidence that such an effect exists is not definitive,
but is plausible given that correcting for a single local or global mass
step leaves an additional step with $\sim 3 \sigma$ significance.
In a sample of SNe\,Ia for which global and local indicators
disagree, we see \textit{no evidence} for a local or global step
as a function of either mass or color.
Figure \ref{fig:localvglobal}
summarized the Hubble residuals in each local versus global quadrant.
We find 1.7$\sigma$ evidence for a local $u-g$ step after
correcting for a global host mass step.

Though the results here do not prove that SNe\,Ia are more correlated
with their local host environments than their global environments,
we use these results and their uncertainties
to put limits on the estimated bias to cosmological
parameters due to local effects.  The only step detected
at $>$2$\sigma$ significance, the local mass step, would give an estimated systematic
shift in H$_0$ of -0.14 km s$^{-1}$ Mpc$^{-1}$ with an additional uncertainty
of 0.14 km s$^{-1}$ Mpc$^{-1}$, $\sim$10\% of the current
uncertainty on H$_0$.

Lastly, we find 2.1-2.9$\sigma$ evidence for tension between
measurements of the local step from surveys that target a pre-selected
set of galaxies (the previous low-$z$ sample) and surveys that do not.
Previous work has also shown that different samples may have different step sizes and
it is not clear why (e.g. \citealp{Rest14,Scolnic14}).  \citet{Roman18} found that
the targeted low-$z$ sample has marginal evidence for a local color step of $0.049 \pm 0.046$ mag (1.1$\sigma$
significance), but they found a local step that was nearly twice as large when
including data with 87\% of SNe from
untargeted surveys (7.0$\sigma$ significance).
The fact that the
untargeted surveys here were observed on the Sloan filter system,
while the targeted surveys used Johnson filters may also perhaps
play a role.  Though
the samples included in \citet{Roman18} cannot determine whether this result is due to redshift
evolution of the step or survey-specific effects, our data $-$ and future
Foundation data releases $-$ can break this degeneracy.

We remain agnostic about the reasons for sample-to-sample
differences, but it is clear that pre-selecting galaxies will alter the
demographics of the SN sample and therefore may change the measured relationships
of SNe\,Ia with their hosts.
As most SNe used in the \citet{Riess16} H$_0$ measurement are
from targeted searches, it is unclear whether
it is appropriate to apply a correction to the current H$_0$
analysis if that correction is measured from untargeted samples.
This question is unlikely to be resolved
without a better understanding of the relationships between
SNe\,Ia and their environments.

The existing low-$z$ sample is also subject to significant
calibration uncertainties and selection biases.
A local mass step in particular could be biased by
difference imaging residuals in SN\,Ia photometry.
In Foundation, we have multiple epochs of PS1 3$\pi$ with no SN light
that can be used to test and correct for the possibility of small
difference imaging biases in future work.
When SNfactory \citep{Aldering02} and the Foundation second
data release are publicly available,
these data may reveal correlations that our data are unable to
probe.

As the connection between SN environments and their progenitors
remains unclear, the SN-host relation will remain a possible source
of systematic uncertainty in cosmological analyses for the foreseeable
future.  If future studies find evidence for a relationship
between SN\,Ia corrected magnitudes and their local
environments, we propose that these studies adopt
the methodology presented here to determine the ``locality''
of the correlation.  If global host properties
will be sufficient to correct SN\,Ia magnitudes
for host galaxy biases, space-based
imaging will not be needed for precision cosmology.
If, on the other hand, convincing evidence is shown
that regions 5 kpc from the SN location are
not as well correlated with the SN\,Ia corrected magnitude
as regions 2 kpc from the SN location, this would
have enormous consequences for future cosmological
analyses and the resources such analyses would
require.

\acknowledgements
We would like to thank the anonymous referee for many helpful suggestions
to improve this manuscript.  We would also like to thank Marcin Sawicki, Mickael Rigault,
and Ravi Gupta for their assistance in improving the measurements
and text.
D.O.J. is supported by a Gordon and Betty Moore Foundation
postdoctoral fellowship at the University of California, Santa Cruz.
The UCSC group is supported in part by NASA grant NNG17PX03C, NSF
grant AST-1518052, the Gordon \& Betty Moore Foundation, the
Heising-Simons Foundation, and by fellowships from the Alfred P.\
Sloan Foundation and the David and Lucile Packard Foundation to R.J.F.

\appendix
\section{Intrinsic Dispersion Measurements for each Subsample}

In this appendix, we reproduce Tables \ref{table:masscorr} and \ref{table:random} 
but list dispersion values instead of mass and color step measurements for each
subsample (Tables \ref{table:masscorrdisp} and \ref{table:randomdisp}).
We measure these dispersions using the likelihood model presented in
\S\ref{sec:likemod}.  Occasionally, dispersions are equal to zero, but with high uncertainty
meaning that photometric errors alone appear to explain the scatter about the
Hubble diagram.

We see $\sim 1-2 \sigma$ evidence that SNe\,Ia in locally low-mass or
locally blue regions have lower dispersion.  However, we do not see a significant
difference between the ``local'' and ``random'' measurements for the dispersion.

\begin{deluxetable*}{lrrrrrrr}
\tabletypesize{\scriptsize}
\tablewidth{0.95\columnwidth}
\tablecaption{Measurements of the Hubble Residual Dispersion for Targeted and Non-Targeted Surveys}
\tablehead{&\multicolumn{3}{c}{$\Delta_M$}&&\multicolumn{3}{c}{$\Delta_{u-g}$}\\*[2pt]
\cline{2-4} \cline{6-8} \\*[2pt]
&log(M$_{\ast}$/M$_{\odot}$ $<$ 8.9)&log(M$_{\ast}$/M$_{\odot}$ $>$ 8.9)&diff.&&$u-g < 1.6$&$u-g > 1.6$&diff.}
\startdata
Local Step&$0.093\pm0.010$&$0.118\pm0.011$&$-0.025\pm0.015$&&$0.108\pm0.009$&$0.048\pm0.043$&$0.060\pm0.044$\\
$-$ Targeted SNe&$0.067\pm0.018$&$0.102\pm0.026$&$-0.035\pm0.032$&&$0.084\pm0.016$&$0.000\pm0.063$&$0.084\pm0.065$\\
$-$ No Targeted SNe&$0.101\pm0.013$&$0.128\pm0.015$&$-0.026\pm0.020$&&$0.122\pm0.012$&$0.072\pm0.042$&$0.050\pm0.044$\\
\\
Global Step&$0.106\pm0.013$&$0.111\pm0.009$&$-0.005\pm0.015$&&$0.107\pm0.009$&$0.078\pm0.042$&$0.029\pm0.043$\\
$-$ Targeted SNe&$0.089\pm0.034$&$0.081\pm0.015$&$0.008\pm0.037$&&$0.085\pm0.016$&$0.000\pm0.047$&$0.085\pm0.050$\\
$-$ No Targeted SNe&$0.115\pm0.016$&$0.127\pm0.012$&$-0.012\pm0.020$&&$0.119\pm0.012$&$0.107\pm0.044$&$0.012\pm0.045$\\
\enddata
\tablecomments{Similar to Table \ref{table:masscorr}, except that after correcting for global host galaxy mass
  we give the measurements of SN\,Ia intrinsic dispersion for
  subsamples of SN\,Ia in different local or global environments.  We measure dispersion using free parameters
in the likelihood model presented in \S\ref{sec:likemod}.}
\label{table:masscorrdisp}
\end{deluxetable*}

\begin{deluxetable*}{lrrrrrrr}
\tabletypesize{\scriptsize}
  \tablewidth{0.95\columnwidth}
\tablecaption{Measurements of the Hubble Residual Dispersion after Correcting SN\,Ia for their Local or Random Environments}
\tablehead{&\multicolumn{3}{c}{$\Delta_M$}&&\multicolumn{3}{c}{$\Delta_{u-g}$}\\*[2pt]
\cline{2-4} \cline{6-8} \\*[2pt]
&log(M$_{\ast}$/M$_{\odot}$ $<$ 8.9)&log(M$_{\ast}$/M$_{\odot}$ $>$ 8.9)&diff.&&$u-g < 1.6$&$u-g > 1.6$&diff.}
\rowcolor{gray!25}Local Step&$0.094\pm0.016$&$0.116\pm0.027$&$-0.022\pm0.031$&&$0.089\pm0.015$&$0.115\pm0.014$&$-0.026\pm0.020$\\
\rowcolor{gray!25}Random Step\tablenotemark{a}&$0.084\pm0.009$&$0.127\pm0.012$&$-0.043\pm0.015$&&$0.096\pm0.015$&$0.112\pm0.014$&$-0.016\pm0.021$\\
5 kpc from SNe&$0.101\pm0.012$&$0.119\pm0.016$&$-0.018\pm0.020$&&$0.087\pm0.013$&$0.127\pm0.015$&$-0.040\pm0.020$\\
10 kpc from SNe&$0.110\pm0.012$&$0.114\pm0.012$&$-0.004\pm0.017$&&$0.101\pm0.015$&$0.109\pm0.014$&$-0.009\pm0.021$\\
$R < 1$&$0.094\pm0.011$&$0.110\pm0.012$&$-0.015\pm0.016$&&$0.074\pm0.015$&$0.128\pm0.018$&$-0.055\pm0.024$\\
$1 < R < 2$&$0.101\pm0.010$&$0.120\pm0.012$&$-0.020\pm0.016$&&$0.094\pm0.013$&$0.118\pm0.014$&$-0.024\pm0.019$\\
$2 <  R < 3$&$0.087\pm0.011$&$0.129\pm0.012$&$-0.042\pm0.016$&&$0.092\pm0.015$&$0.118\pm0.014$&$-0.026\pm0.021$\\
\tablenotetext{a}{Regions $>$5 kpc from SN are randomly sampled.  One
  random region is chosen per SN, the step is measured, and this
  process is repeated 100 times.  The steps listed
  here are the mean of 100 samples.}
\tablecomments{$R$ is the distance from the center of the galaxy
  in units of the normalized elliptical radius of the galaxy \citep{Sullivan06}.  The last 5
  rows exclude regions within 3 kpc of the SN location.  Also in the last 5 rows,
  the step location is taken to be the median of every sample to avoid a situation in which
  90\% or more of the sample is considered ``high-mass'' or ``low-mass''.}
\tablecomments{Similar to Table \ref{table:random}, except that but we give the measurements
  of SN\,Ia intrinsic dispersion for subsamples of SN\,Ia using the likelihood model
  presented in \S\ref{sec:likemod} (after correcting for host galaxy mass).  We
  explore how the difference in dispersion between samples of
  SN\,Ia with different host characteristics evolves when SN\,Ia properties are inferred
  from random regions or regions far from the SN location.}
\label{table:randomdisp}
\end{deluxetable*}

\bibliographystyle{apj}

\end{document}